\documentclass[prd,aps,showpacs,groupedaddress,eqsecnum,notitlepage,nofootinbib,superscriptaddress]{revtex4-2}
\usepackage{graphicx}

\usepackage{tikz,siunitx,mwe}
\usepackage{tensor}
\usepackage{epstopdf}
\usepackage{amsmath}
\usepackage{amsfonts}
\usepackage{slashed}
\usepackage{amssymb}
\usepackage{enumerate, color}
\usepackage{subfigure}
\usepackage{lipsum}
\usepackage{color}
\usepackage{graphicx,bm}
\usepackage[utf8]{inputenc}
\usepackage{eurosym}
\usepackage{scalerel}%To smaller subindex
\usepackage{float}
\usepackage{accents}
\usepackage[colorlinks=true,pdfstartview=FitV,linkcolor=blue,citecolor=blue,urlcolor=blue,breaklinks=true]{hyperref}
\usepackage{orcidlink}
\newcommand{\be}{\begin{equation}}
\newcommand{\ee}{\end{equation}}
\newcommand{\ben}{\begin{eqnarray}}
\newcommand{\een}{\end{eqnarray}}
\newcommand{\bes}{\begin{subequations}}
\newcommand{\ees}{\end{subequations}}
\usepackage{comment}
\def\bal#1\eal{\begin{align}#1\end{align}}

\def\bal#1\eal{\begin{align}#1\end{align}}
\begin{document}
%\title{ The scattering of massless quasiparticles in graphene with disclinations}
\title{Contribution of Geometry and Non-Abelian Gauge Fields to Aharonov-Bohm Scattering of Massless Fermions in Graphene with Disclinations}

%%%%%%%%%%%%%%%%%%%%%%%%%%%%%%%%%%%%%%%%%%%%%%%%%%%%%%%%%%%%%%%%%%%%%%%%%%%%%%

%-------------------------------------------%
\author{M. J. Bueno\,\orcidlink{0000-0001-7588-7150}}
\email{jannaira@gmail.com}
\affiliation{Departamento de F\'isica, Universidade Federal da Para\'iba, 58051-970, Jo\~ao Pessoa, PB, Brazil. }
%-------------------------------------------%
%-------------------------------------------%
\author{G. Q. Garcia\,\orcidlink{0000-0003-3562-0317}}
\email{gqgarcia99@gmail.com}
\affiliation{Centro de Ci\^encias, Tecnologia e Sa\'ude, Universidade Estadual da Para\'iba, 58233-000, Araruna, PB, Brazil.}
%-------------------------------------------%

%-------------------------------------------%
\author{A. M. de M. Carvalho\,\orcidlink{0009-0006-3540-0364}}
\email{alexandre@fis.ufal.br}
\affiliation{Instituto de F\'{\i}sica, Universidade Federal de Alagoas, 57072-970, Macei\'o,  AL, Brazil.}
%-------------------------------------------%
\author{C. Furtado\,\orcidlink{0000-0002-3455-4285}}
\email{furtado@fisica.ufpb.br}
\affiliation{Departamento de F\'isica, Universidade Federal da Para\'iba, 58051-970, Jo\~ao Pessoa, PB, Brazil. }
%-------------------------------------------%
%-------------------------------------------%
%%%%%%%%%%%%%%%%%%%%%%%%%%%%%%%%%%%%%%%%%%%%%%%%%%%%%%%%%%%%%%%%%%%%%%%%%%%%%%
\begin{abstract}
This work examines the effect of disclinations on the scattering of quasipaticles in graphene with the presence of a topological defect. Using the tight-binding method, the electronic properties of graphene with disclination are described, where the  topological defects are introduced in the lattice via geometric theory. The massless Dirac equation is modified to account for the curvature induced by these defects, incorporating a gauge field. The results show that disclinations significantly affect the scattering process, altering phase shifts and interference patterns. The differential cross-section and its dependence on the scattering angle are analyzed, highlighting the role of geometric factors like the parameter \(\alpha\) in shaping the scattering dynamics.
\end{abstract}
%%%%%%%%%%%%%%%%%%%%%%%%%%%%%%%%%%%%%%%%%%%%%%%%%%%%%%%%%%%%%%%%%%%%%%%%%%%%%%
%%%%%%%%%%%%%%%%%%%%%%%%%%%%%%%%%%%%%%%%%%%%%%%%%%%%%%%%%%%%%%%%%%%%%%%%%%%%%%
\maketitle
%%%%%%%%%%%%%%%%%%%%%%%%%%%%%%%%%%%%%%%%%%%%%%%%%%%%%%%%%%%%%%%%%%%%%%%%%%%%%%
\section{Introduction}

The study of graphene has generated significant interest due to its remarkable electronic properties, such as its high electrical conductivity and mechanical strength, which make it a promising material for a range of applications in nanotechnology and quantum devices~\cite{Novoselov2005PNAS,geim2007rise,Novoselov2005Nature}. These properties arise from its unique honeycomb lattice structure, which, under certain conditions, can exhibit massless Dirac fermions at low energies. Because of this, graphene sheet attracted attention as a material where field theory can be studied in a practical way. In 1992, Katanaev and Volovich~\cite{katanaev1992} established a geometric theory of defects in solids, relating torsion and curvature in elastic media to topological defects in the lattice. This work laid the foundation for understanding how disclinations could be treated as geometric singularities, which, in the context of graphene, could be modeled using the Dirac equation in curved spaces. The resulting effects on the electronic properties of graphene have been studied using these geometric frameworks~\cite{katanaev1992}. Therefore, the geometric theory of defects made graphene an excellent analogous gravitational model in condensed matter physics.

Since the discovery of graphene, various studies have focused on understanding how the presence of topological defects, such as disclinations, can influence its electronic properties. Disclinations, which are topological defects resulting from the insertion or removal of angular sectors in a lattice, introduce local curvature in the material~\cite{volterra1907equilibre}. In graphene, these defects are typically associated with the formation of pentagon or heptagon rings, leading to changes in the lattice symmetry and affecting the scattering of quasiparticles~\cite{vozmediano2010gauge, pachos2009manifestations}. These topological defects can transform a flat graphene sheet into curved structures, such as graphitic cones~\cite{furtado2008, oliveira2017evolution, oliveira2021graphene}, fullerene~\cite{gonzalez1993electronic, PhysRevLett.69.172}, graphene wormhole~\cite{Gonzalez, Garcia:2019gro, garcia2025rotation}, and so on. Subsequent studies, such as the work on the low-energy electronic spectrum of graphene in the presence of disclinations, explored the effects of an external magnetic field. Using a continuum approach, it was demonstrated that disclinations modify the Landau levels significantly, with the energy spectrum explicitly depending on the disclination parameter and the magnetic field~\cite{bueno2012}. In 2008, a study used a geometric approach to analyze the impact of disclinations in graphitic cones. This study showed that a spinor, which describes low-energy states near the Fermi level, acquires a phase when transported around the apex of the cone. This result is directly due to the topological defect, and the phase acquisition is analogous to the Aharonov-Bohm effect. The study extended the analysis to systems with multiple cones, providing a comprehensive description of how disclinations in graphene can lead to non-trivial geometric phases and affect the electronic properties of the material~\cite{furtado2008}. Recently, Fernandez {\it et al.} ~\cite{pujol} have investigated the electronic properties of graphene with ddisclination using a geometric theory of defects.
Holonomic Quantum Computation in graphene sheets with disclination was investigated using geometric theory of defects in ~\cite{kolonomygraphene}. One of Us\cite{bakke2012kaluza} have investigate a geometric phase in graphene with disclination emploing a Kaluza-Klein theory to describe a elatic media with defects.

Our work is focused on an analysis of the quasiparticle scattering in the graphitic cone employed by the Deser and Jackiw method~\cite{Deser1988} in the investigation of solutions of scattering of particle in $(2+1$-gravitational background of poin particle. In their work, Deser and Jackiw studied the scattering of a particle on a conical background geometry, where a partial expansion of the wavefunction solutions in a cone was used. They have considered the absence of potential term, and geometry information was given by the kinetic term. Several papers on the transport properties in graphene have been published in the last few decades. Ando {\it et al.} reported the lack of back scattering in a carbon nanotube due to the Berry phase acquired by wave vector space rotation~\cite{ando1998berry}. In 2009, an experiment was conducted to measure the effects of defects in charge transport in graphene~\cite{chen2009defect}. The study of particle scattering on the surface of the graphene sheet for posite and negative cone\cite{fonseca2010scattering} In this experiment, Chen {\it et al.} showed that defects have a significant effect on charge scattering. In ref.~\cite{vaishnav2011intravalley} the researchers developed a theoretical description for the intravalley scattering of quasiparticles on the graphene by Green's function method. Other experimental references on this subject can be found in ~\cite{rutter2007scattering, bostwick2007quasiparticle}. 

Our study on the scattering of quasiparticles in graphene with disclinations presents conceptual and methodological differences compared to previous works, such as that of Fonseca et al \cite{fonseca2010scattering}. In our approach, in addition to curvature, we introduce a non-Abelian gauge field, reflecting the interaction between quasiparticles and the electronic structure altered by the disclination. As a consequence, our results show that the differential cross-section and the phase shifts in the scattering process are influenced not only by the defect's curvature but also by the gauge structure that emerges from the lattice deformation. Therefore, our analysis provides a more comprehensive model for understanding how disclinations affect electronic scattering in graphene, highlighting the importance of gauge effects in modifying the electronic transport properties of two-dimensional materials.

This paper is organized as follows: In Section~\ref{sec2}, we describe graphene in a curved background due to the presence of a disclination. In Section~\ref{sec3}, we obtain the solutions of the massless Dirac equation in terms of Bessel functions. In Section~\ref{sec4}, we analyze the quasiparticle scattering in graphene using the Deser and Jackiw method. In Section~\ref{sec5}, we present the results and discussion, highlighting the effects of disclinations on the differential cross-section and interference patterns. Finally, in Section~\ref{sec6}, we summarize our conclusions and discuss future perspectives.

%%%%%%%%%%%%%%%%%%%%%%%%%%%%%%%%%%%%%%%%%%%%%%%%%%%%%%%%%%%%%%%%%%%%%%%%%%%%%%
\section{Description of graphene in the presence of disclination}\label{sec2}

Our main goal is to study the influence of disclinations on the scattering of graphene quasiparticles. For this, we need to describe the graphene quasiparticle considering a curved background with disclinations. The graphene monolayer is a carbon honeycomb lattice described by superposition of two triangular sublattices $\mathcal{A/B}$ \cite{geim2007rise, katsnelson2007graphene}. The carbon atom in the lattice site is bonded with its three nearest neighbors through $\sigma$ bonds, and its last valence electron forms a $\pi$-bond. These $\sigma$ bonds give us the elastic properties of the lattice, and the $\pi$-bond is responsible for the electronic properties. Our focus is on the electronic properties so that we can describe it using the tight-binding (TB) method. In this method, we are considering the valence electron in the $\pi$-bond of carbon site hops for one of the three nearest neighbors. In the literature \cite{katsnelson2007graphene, vozmediano2010gauge, Garcia:2019gro} the TB method for graphene was widely solved and, as a consequence, a dispersion relation described by two Fermi points $K/K'$ was obtained. For low moments, the dispersion relation is linear, and the quasiparticle on graphene has a relativistic behave. Therefore, we can use the massless Dirac Hamiltonian in $(2+1)$-dimensions to obtain the graphene electronic properties:
\begin{equation}
    H = -i\hslash v_f \Vec{\sigma}\cdot\Vec{\nabla},
    \label{HMDirac}
\end{equation}
where $v_f$ is the Fermi velocity and $\vec{\sigma}= \lbrace{\sigma_1,\sigma_2\rbrace}$ is a set of the Pauli matrices. To introduce a curved background in eq. \eqref{HMDirac}, we must substitute the momentum operator $\vec{\nabla}$ for the covariant derivative operator $\nabla_{\mu} = \partial_{\mu} - \Gamma_{\mu}$. Thus, the massless Dirac equation can be rewrite as follow:
\be
    i\gamma^{\mu}(\partial_{\mu} - \Gamma_{\mu} - i\Omega_{\mu})\Psi = 0.
    \label{EMDirac}
\ee
We see that $\Gamma_\mu$ is the spinnorial connection responsible for the curved character of the massless Dirac equation and $\Psi$ is the $2$-component spinnor. Beyond the introduction of a curved space background, the expression for the massless Dirac equation \eqref{EMDirac} is also useful for the insertion of topological defects through the gauge field $\Omega_\mu$.

One of the most common topological defects in crystals is disclination. In graphene, disclinations are characterized by the presence of pentagon or heptagon rings in its lattice. Disclinations was described by Volterra in ref. \cite{volterra1907equilibre} through of the ``cut and glue" processes, where a angular sector of material is removed (or inserted) and its borders are glued together. As graphene is a material with $(2+1)$-dimensions, after this ``cut and glue" process, it assumes a curved background with a pentagon or heptagon ring in the core of the topological defect. For a pentagon ring, the graphene assumes a positive curvature, whereas for a heptagon ring, it assumes a negative curvature. By the geometric theory of defects in solids \cite{katanaev1992theory, katanaev2023combined}, disclinations are related to spacetime curvature. Another consequence of the presence of these types of topological defect in the graphene lattice is the connection between sites of the same kind. In other words, the presence of disclination mixes the Fermi point $K/K'$ in the momentum space \cite{pachos2009manifestations,bueno2012landau}, and therefore it is necessary to introduce a gauge field $SU(2)$ to compensate for this graphene distortion. Thus, 
\be
    \oint \Omega_{\phi} d\phi = -3\pi (\alpha-1) \tau_y,
    \label{SU(2)GF}
\ee
where $\tau$ is the set of Pauli matrices in the reciprocal space and the parameter $\alpha$ is related to the angular sector in the ``cut and glue" process \cite{pachos2009manifestations, bueno2012landau}.

As mentioned previously, graphene in the presence of a dislocation presents a non-null curvature tensor. In fact, the curved background of graphene with dislocation can be written as the following line element \cite{bueno2012landau}: 
\be
    ds^{2}=dt^{2}-d\rho^{2}-\alpha^{2}\rho^{2}d\phi^{2}.
    \label{linelement}
\ee
This line element is known as the conical spacetime, where the parameter $\alpha$ gives us information about the removal (or insertion) of the angular sector in the formation of conical defects. This geometry has non-null curvature tensor only on the defect core, and this behavior was called a conical singularity. When the $\alpha$ parameter assumes values between zero and 1 $(0<\alpha<1)$, the angular sector was removed from the plane and a pentagon ring rises in the core defect. In this case, the curvature is positive and the geometry is a cone. On the other hand, when the parameter $\alpha$ assumes values greater than 1 $(\alpha>1)$, a material angular sector is inserted so that a heptagon ring rises in the core defect and its geometry has a negative curvature. In order to describe the massless Dirac equation in this curved background \eqref{linelement}, we adopt the Maurer-Cartan formalism \cite{nakahara2018geometry}. In this formalism, we can define the non-coordinate basis $\hat{\theta}^{a} = e^{a}_{\ \mu} dx^{\mu}$, where $e^{a}_{\ \mu}(x)$ is called {\it vielbien} and the line element can be rewritten in terms of the Minkowski tensor $ds^2 = \eta_{ab} \hat{\theta}^{a} \hat{\theta}^{b}$. Thus, {\it vielbien} and its inverse can be defined as
\ben
    e^{a}_{\ \mu}(x) = \left(\begin{array}{ccc}
       1  &           0           &          0            \\
       0  &      \cos{\phi}       & -\alpha\rho\sin{\phi} \\
       0  &      \sin{\phi}       &  \alpha\rho\cos{\phi}
    \end{array}\right)\ \ \text{and}\ \ e_{a}^{\ \mu}(x) = \left(\begin{array}{ccc}
       1  &             0                &               0             \\
       0  &         \cos\phi             &           \sin\phi          \\
       0  & -\frac{\sin\phi}{\alpha\rho} &  \frac{\cos\phi}{\alpha\rho}
    \end{array} \right).
    \label{vielbien}
\een
Solving the Maurer-Cartan structures $T^a = d\hat{\theta}^{a} + \omega^{a}_{\ b}\wedge\hat{\theta}^{b}$, considering the torsion tensor $T^a = 0$, we obtain the non-vanishing spin connection components $\omega^{a}_{\ b} = \omega^{\ a}_{\mu\ b} dx^{\mu}$ as follow
\be
    \omega^{\ 1}_{\phi\ 2} = -\omega^{\ 2}_{\phi\ 1} = (\alpha - 1),
    \label{spinconnect}
\ee
that is related with the spinnorial connection $\Gamma_{\mu} = \frac{1}{8}\omega_{\mu ab} \left[\sigma^a,\sigma^b\right]$.
%%%%%%%%%%%%%%%%%%%%%%%%%%%%%%%%%%%%%%%%%%%%%%%%%%%%%%%%%%%%%%%%%%%%%%%%%%%%%%
\section{Massless Dirac equation for graphene with disclination}\label{sec3}

In order to obtain the quantum dynamics of the graphene quasiparticles, we must introduce the curved aspects, previously presented in Section \ref{sec2}, in the massless Dirac equation. Thus, we can rewrite the massless Dirac equation \eqref{EMDirac} in the disclination background through the assistance of curved space elements \eqref{vielbien} and \eqref{spinconnect}, and introducing a non-Abelian gauge field \eqref{SU(2)GF}. Consequently we have obtained the follow expression:
\be
    i\gamma^t \frac{\partial\Psi}{\partial t} + i\gamma^{\rho}\left[\frac{\partial}{\partial \rho} + \frac{(\alpha - 1)}{2\alpha\rho}\right]\Psi + \frac{i\gamma^{\phi}}{\alpha\rho} \left[\frac{\partial}{\partial \phi} + i\alpha\Omega_{\phi}\right]\Psi = 0.
    \label{MDequation2}
\ee
The set of gamma matrices is defined in terms of Pauli matrices $\sigma^a$ by the expression $\gamma^{\mu} = e_{a}^{\ \mu} \sigma^{a}$. In particular, we can define the gamma matrices as a result of the above equation,
\bes
\ben
    \gamma^t &=& \sigma^0 = \mathbf{I}_{2\times 2};\\
    \gamma^{\rho} &=& \cos{\phi}\ \sigma^1 + \sin{\phi}\ \sigma^2;\\
    \gamma^{\phi} &=& \frac{1}{\alpha\rho}\left(\cos{\phi}\ \sigma^2 - \sin{\phi}\ \sigma^1 \right).
\een    
\ees
Note that the massless Dirac equation is a two-component coupled differential equation. With the purpose to decouple the Dirac spinor we realize a rotation on the plane \cite{villalba1994exact, villalba2001energy}. This rotation can be represented by a similarity transformation in which the gamma matrices $\gamma^{\rho}$ and $\gamma^{\phi}$ can be turned in the Pauli matrices $\sigma^1$ and $\sigma^2$, respectively. Thus, this similarity transformation is written as
\be
    S(\phi) = exp\left[-\frac{i\phi}{2}\sigma^{3}\right],
\ee
obeying the following relations:
\bes
\ben
S^{-1}(\phi)\gamma^{\rho}S(\phi) &=& \sigma^1;\\
S^{-1}(\phi)\gamma^{\phi}S(\phi) &=& \sigma^2.
\een
\ees

After performing this similarity transformation, the Dirac spinor is given by $\Psi = S(\phi)\Bar{\Psi}$ and the massless Dirac equation \eqref{MDequation2} assumes the form below
\be
    i\frac{\partial\Bar{\Psi}}{\partial t} + i\sigma^{1}\left(\frac{\partial}{\partial\rho} + \frac{1}{2\rho}\right)\Bar{\Psi} + \frac{i\sigma^{2}}{\alpha\rho} \left(\frac{\partial}{\partial\phi} + i\alpha\Omega_{\phi}\right)\Bar{\Psi} = 0.
    \label{MDequation3}
\ee
One method to solve this massless Dirac equation \eqref{MDequation3} is to choose a {\it ansatz} for the spinor in the following way
\ben
    \Bar{\Psi} = e^{-iEt+i\left(m+\frac{1}{2}\right)\phi} \left(\begin{array}{c}
         \Bar{\psi}_{A}  \\
         \Bar{\psi}_{B}
    \end{array} \right),
\een
so that we obtained a system with two coupled first-order differential equations. Therefore, the equation \eqref{MDequation3} can be written in a matrix form:
\ben
    \left(\begin{array}{cc}
       E  & -i\left[\frac{d}{d\rho} + \frac{1}{2\rho} +\frac{m+\frac{1}{2}}{\alpha\rho} + \frac{\Omega_{\phi}}{\rho}\right] \\
     -i\left[\frac{d}{d\rho} + \frac{1}{2\rho} - \frac{m+\frac{1}{2}}{\alpha\rho} - \frac{\Omega_{\phi}}{\rho}\right]  &  E\end{array}\right)  \left(\begin{array}{c}
         \Bar{\psi}_{A}  \\
         \Bar{\psi}_{B}
    \end{array} \right) = 0.
\een

Finally, by decoupling these two differential equations using the substitution method, we have obtained a decoupled set of two differential equations of the second order. Then we found
\be
    \rho^{2}\frac{d^{2}\Bar{\psi}_{s}}{d\rho^{2}} + \rho \frac{d\Bar{\psi}_{s}}{d\rho} + \left(E^{2}\rho^{2} - \mu^{2}_{s}\right)\Bar{\psi}_{s} = 0,
\ee
where the index $s = A,B$ indicates the degrees of freedom of the sublattice space. The parameters $\mu_s$, with $\sigma_z = \pm 1$ for each sublattice, are given by 
\be
    \mu^{2}_{s} = \left(\frac{m+\frac{1}{2}}{\alpha} + \Omega_{\phi} - \frac{\sigma_z}{2}\right)^{2}.
\ee
Realizing the coordinate transformation $\rho = x/E$, we obtain the Bessel differential equation:
\be
    x^2 \frac{d^{2}\Bar{\psi}_{s}}{dx^2} + x \frac{d\Bar{\psi}_{s}}{dx} + (x - \mu^{2}_{s}) \Bar{\psi}_{s} = 0.
    \label{Bessel}
\ee
This Bessel differential equation has a general solution \cite{machado2012equaccoes} in the following way
\be
    \Bar{\psi}_{s}\left(\dfrac{x}{E}\right) = A J_{\mu_{s}}(x) + B Y_{\mu_{s}}(x),
    \label{solutions}
\ee
Here we see that the parameters $A$ and $B$ are constants, and $J_{\mu_{s}}(x)$ and $Y_{\mu_{s}}(x)$ are Bessel functions of the first and second kind, respectively. Note that the Bessel functions of the first kind $J_{\mu_{s}}(x)$ are finite at the origin, while the Bessel functions of the second kind are divergent at the origin. In order for the wave function to behave well at the origin, we choose the constant $B$ as being null, and the solution is given by Bessel functions of the first kind only.

%%%%%%%%%%%%%%%%%%%%%%%%%%%%%%%%%%%%%%%%%%%%%%%%%%%%%%%%%%%%%%%%%%%%%%%%%%%%%%
\section{Quasiparticle Scattering on the Graphene}\label{sec4}

For now, we analyze the scattering of quasiparticles on graphene using the method of Deser and Jackiw~\cite{Deser1988}. For this purpose, we take the solutions~\eqref{solutions} of the Dirac equation to study the relative scattering of the Dirac fermions. Analyzing the temporal evolution of the motion of a particle incident from the remote past ($t \to -\infty$) toward the future ($t \to \infty$), we consider the scattering of the incident plane wave by a structured defect in graphene. Redefining the parameters as $\mu_{A} = \nu$ and $\mu_{B} = \nu+1$, and returning to the original coordinate $\rho$, we can build the quasiparticle scattering wavefunction by partial wave expansion in terms of the Dirac spinor~\eqref{solutions} as pointed out in ref~\cite{Deser1988}. Therefore,
\begin{eqnarray}\label{super}
\psi(\rho,\phi) &=&\sum_{l=-\infty}^{\infty} i^{l}e^{i\delta_{l}}
 \,\, 
\left(
\begin{array}{cc}
\Bar{\psi}^{A}_{l}(\rho)\\
\Bar{\psi}^{B}_{l+1}(\rho)\\
\end{array}\right)\,\,e^{il\phi} ,
\end{eqnarray}
where $i^l= e^{il\frac{\pi}{2}}$ and using the following expansion for the plane wave function: $e^{i E\rho \cos \phi}= \sum_{l=-\infty}^{\infty} i^{l} J_{l}(\rho)e^{il\phi}$. If we assume that the interaction region is spatially restrict, we can consider the asymptotic limit where admissible solutions for the wave function take the form:
\begin{subequations} 
\begin{eqnarray}
\lim_{\rho \to \infty} \Bar{\psi}_{l}\left(\rho\right) &=& \sqrt{\frac{2}{\pi E\rho}} \cos\left(E \rho -\frac{l\pi}{2} - \frac{\pi}{4} + \delta^{A}_{l} \right);\label{bessel1}\\ 
\lim_{\rho \to \infty} \Bar{\psi}_{l +1}\left( \rho \right) &=& \sqrt{\frac{2}{\pi E\rho}} \cos\left(E \rho - \frac{\left(  l + 1 \right) \pi}{2} - \frac{\pi}{4} + \delta^{B}_{l+1} \right).\label{bessel2}
\end{eqnarray}
\end{subequations}
Note that $\delta^{A}_{l}$ and $\delta^{B}_{l}$ are the phase shift due to sublattices $\mathcal{A}$ and $\mathcal{B}$, respectively. And to determine the phase shift ($\delta_l$), we equate the arguments of equations \eqref{bessel1} and \eqref{bessel2}. Defining $\omega=\left(\alpha^{-1} - 1 \right)\pi$, we have obtained that
\begin{eqnarray}
\delta_{l} = \delta^{A}_{l} = \delta^{B}_{l+1}= - \left[\left( l + \frac{1}{2}\right) \frac{\omega}{2} + \Omega_{\phi}\frac{\pi}{2}\right].
\end{eqnarray}
We can conclude that the phase shift angle, $\delta_{l}$, does not depend only on the influence of the geometry of the defect contained in $\omega$, as observed by Fonseca {\it et al.}~\cite{fonseca2010scattering}, but also depends on $\Omega_{\phi}$ which is related to the presence of the non-Abelian gauge field. Therefore, we have two contributions: one related to the Aharonov-Bohm effect due to the geometry, and the second contribution due to the non-Abelian gauge field. The phase shift can also be expressed in terms of the number of removed sectors $N$. Its dependence on $N$ reflects the angular deformation caused by the disclination in graphene, which modifies the curvature of the effective space where Dirac fermions propagate, affecting their dynamics. The term $N/(6-N)$ 
quantifies the angular deficit introduced by the defect, indicating how the removal of sectors changes the boundary conditions and alters the effective angular momentum of the quantum states. As a result, the phase of the electronic states is adjusted, directly influencing the scattering of the quasiparticles
\begin{eqnarray}\label{delta3}
\delta_{l} = \left[ - \left( l + \frac{1}{2}\right) \frac{N}{6-N}\pi \pm \frac{N\pi}{8}\right].
\end{eqnarray}
The table \ref{tabela} presents the relationship between $N$, $\alpha$, and $\Omega$. The parameter $\alpha$ defines the effective curvature of the disclination cone, with $\alpha < 1$ for sector removal and $\alpha > 1$ for sector addition. The gauge field $\Omega = \pm \tfrac{3}{2}(\alpha - 1)$ emerges from the sublattice mixing and Dirac point shifts, with the sign determined by the Dirac valley. For example, when $N = 2$, $\alpha = \tfrac{2}{3}$, yielding $\alpha - 1 = -\tfrac{1}{3}$ and $\Omega = \mp\tfrac{1}{2}$. Thus, $N$ and $\alpha$ dictate the geometric deformation, while $\Omega$ captures the gauge contribution, together fully characterizing the scattering behavior in graphene with disclinations. There are four kinds of defects for the $N$ values \cite{choudhari2014graphene}: For $N=1$ the disclination is a pentagon defect, $N=2$ is the square defect, $N=-1$ is the heptagon defect and $N=-2$ is the octagon defect.
%%%%%%%%%%%%%%%%%%%%%%%%%%%%%%
\begin{table}[ht]
\centering
\caption{Values of $N$, $\alpha$, and $\Omega$ from $\alpha = 1 - \tfrac{N}{6}$ and $\Omega = \pm\tfrac{3}{2}(\alpha - 1)$.}
\vspace{1em}
\renewcommand{\arraystretch}{1.5} % Increase row spacing
\setlength{\tabcolsep}{12pt}      % Increase column spacing
\begin{tabular}{|c|c|c|c|}
\hline
$\mathbf{N}$ & $\mathbf{\alpha}$ & $\mathbf{\Omega}$ & Type of defect\\
\hline
-2 & $\tfrac{4}{3}$ & $\pm\tfrac{1}{2}$ & Octagon defect\\
\hline
-1 & $\tfrac{7}{6}$ & $\pm\tfrac{1}{4}$ & Heptagon defect\\
\hline
0  & 1             & 0               & -\\
\hline
1  & $\tfrac{5}{6}$ & $\mp\tfrac{1}{4}$ & Pentagon defect\\
\hline
2  & $\tfrac{2}{3}$ & $\mp\tfrac{1}{2}$ & Square defect\\
\hline
\end{tabular}
\label{tabela}
\end{table}

%%%%%%%%%%%%%%%%%%%%%%%%%%
The scattered wave function $\psi(\rho,\phi)$ can be written as the sum of the incident wave function $\psi_{inc}$ and the scattered wave function $\psi_{sc}$, that is
\begin{eqnarray}
\psi(\rho,\phi) = \psi_{inc}(\rho,\phi) + \psi_{sc}(\rho,\phi).
\end{eqnarray}
In the asymptotic limit, when $\rho\rightarrow\infty$, the behavior for each spinor component is described as:
\bes
\ben
\psi^{A}(\rho,\phi) &\longrightarrow& \sum_{l=-\infty}^{\infty} i^{l} \Bar{\psi}^{A}_l(\rho) e^{il\phi} +  \sqrt{\frac{1}{\rho}} f_{A}(\phi) e^{iE\rho};\label{4.5a}\\
\psi^{B}(\rho,\phi) &\longrightarrow& \sum_{l=-\infty}^{\infty} i^{l} \Bar{\psi}^{B}_l(\rho) e^{il\phi} +  \sqrt{\frac{1}{\rho}} f_{B}(\phi) e^{iE\rho}.\label{4.5b}
\een
\ees
We have that $f_{\mathcal{A}}(\phi)$ is the scattering amplitude for graphene quasiparticle in $\mathcal{A}$ sublattice, and can be express by comparing the equations~\eqref{4.5a} and~\eqref{bessel1}. Thus,
\begin{eqnarray}
    f_{A}(\phi)= \frac{1}{\sqrt{E 2 \pi}}\sum_{l=-\infty}^{\infty}  
    \left(e^{2\delta_{l}} -1 \right)e^{i\left( l\phi-\frac{\pi}{4}\right)}.
\end{eqnarray}
Similarly, for graphene quasiparticle in  $\mathcal{B}$ sublattice, the scattering amplitude takes the form:
\begin{eqnarray}
    f_{\mathcal{B}}(\phi)= \frac{1}{\sqrt{E 2 \pi}}\sum_{l=-\infty}^{\infty}  
    \left(e^{2\delta_{l}} -1 \right)e^{i\left( l\phi+\frac{\pi}{4}\right)} e^{i\phi}.
\end{eqnarray}
The study of scattering critically depends on the asymptotic limit, as it is in this regime that the interaction between the incident wave and the scattering medium is defined. The acquired phase shift $\delta_l$ and the scattering amplitude $f(\phi)$ are fundamental quantities that can only be extracted by analyzing the wavefunction behavior at large values of $\rho$. Thus, the choice of this limit is both a mathematical necessity and a physical requirement to properly define a well-behaved scattering process.

\begin{eqnarray}\label{AmpEsp}
f_{\sigma}(\phi)= \frac{1}{\sqrt{ 2 \pi E}}\sum_{l=-\infty}^{\infty}  
\left(e^{- 2i\left( l + \frac{1}{2}\right) i\frac{N}{6-N}\pi \pm \frac{N\pi}{4}} -1 \right)e^{i\left( l\phi-\frac{\sigma\pi}{4}\right)},
\end{eqnarray}
With $\sigma = \pm 1$, where $+1$ corresponds to sublattice $\mathcal{A}$ and $-1$ corresponds to sublattice $\mathcal{B}$ in the representation for graphene. The total scattering cross-section can be determined by integrating the differential cross-section $d\sigma/d\phi = |f(\phi)|^2$ over the azimuthal angle:
\begin{eqnarray}
\sigma_{\text{tot}} = \int_0^{2\pi\alpha} \left| f(\phi) \right|^2 d\phi.
\end{eqnarray}
 The study of scattering critically depends on the asymptotic limit, as it is in this regime that the interaction between the incident wave and the scattering medium is defined. The acquired phase shift ($\delta_l$) and the scattering amplitude ($f(\phi)$) are fundamental quantities that can only be extracted by analyzing the wavefunction behavior at large values of $\rho$. Thus, the choice of this limit is both a mathematical necessity and a physical requirement to properly define a well-behaved scattering process~\cite{Deser1988}.

\section{Results and Discussion}\label{sec5}
In this section, we analyze the impact of geometry and non-Abelian gauge fields on the scattering of quasiparticles in graphene with disclinations. We study how different values of the parameter $\alpha$, which controls the curvature of the effective space, and the parameter $\Omega$, associated with the $SU(2)$ gauge field induced by the distortion of the crystalline lattice, affect the differential cross-section. Furthermore, we discuss the resulting interference patterns and their implications for electronic transport in graphene.

\begin{figure}[t!]
\centering 
\begin{tabular}{cc}    %%% not \center
\subfigure[For the angular deficit case $0 < \alpha < 1$. The confined angular range results in a higher frequency of oscillations, indicating strong interference effects and localized scattering. Parameters: $N = 1$, $\alpha = 5/6$, and $\Omega = -1/4$.]
{\includegraphics[width=70mm]{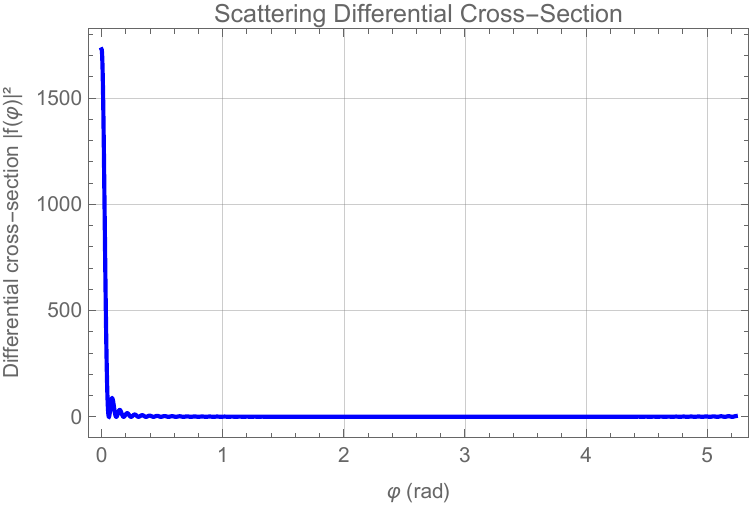}}
\subfigure[For the angular excess case $\alpha > 1$. The extended angular range produces a broader oscillation distribution, decreasing interference density and dispersing scattering intensity over a wider angular region. This result corresponds to $N = -1$, $\alpha = 7/6$, and $\Omega = -1/2$.]
{\includegraphics[width=70mm]{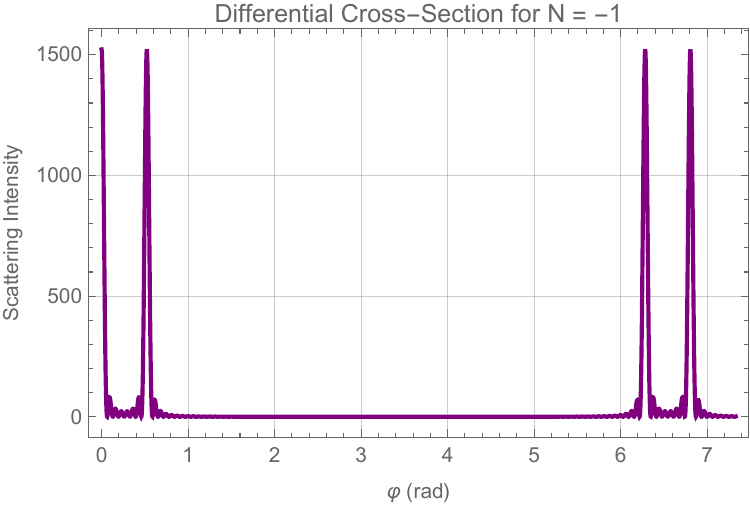}}
\end{tabular}
\caption{Differential cross-section as a function of the scattering angle $\varphi$.}\label{fig1}
\end{figure}

The differential cross-section as a function of the scattering angle exhibits distinct behaviors depending on whether $ \alpha $ is less than or greater than $1$, directly influencing the geometry of the scattering space and modifying the interference structure, as shown in Fig.~(\ref{fig1}). When $ 0 < \alpha < 1 $, the available scattering space is compressed, leading to a reduced angular range $ 2\pi\alpha $, which confines the possible scattering directions. This restriction affects the interference pattern by increasing the density of oscillations in the differential cross-section, as multiple scattered waves interact over a smaller angular domain. The confinement effect also modifies the available scattering modes, leading to a more localized and intensified interference structure where constructive and destructive interferences occur at a higher frequency.

Conversely, when $ \alpha > 1 $, the scattering space is expanded, increasing the angular range and allowing the scattered waves to spread over a larger region. This results in a broader distribution of interference effects, leading to wider and more separated oscillations in the differential cross-section. The increased spatial freedom reduces the density of resonances, making the scattering pattern less localized and more evenly spread across different angles. The key distinction between these two cases is that, in the angular deficit regime ($ 0 < \alpha < 1 $), the system exhibits a higher concentration of resonances, with sharp and closely spaced oscillations, while in the angular excess regime ($ \alpha > 1 $), the interference patterns become more diffuse, with oscillations occurring at larger angular intervals.

Additionally, the vertical axis values in Fig.~(\ref{fig1}) have been adjusted to emphasize the peaks of the differential cross-section, ensuring a clearer visualization of the scattering intensity variations. This adjustment allows for a more accurate comparison of the interference effects across different values of $ \alpha $, highlighting how the geometric control factor dictates whether the scattering process favors localized, high-frequency oscillations or broader, more dispersed interference effects. Ultimately, this influences how energy and wave amplitude are redistributed across different angular directions, reinforcing the role of topology in modifying the scattering behavior.

%%%%%%%%%%%%%%%%%%%%%%%%%%%%%

%%%%%%%%%%%%%%%%%%%%%%%%%%%%%%%%%%%%%%%

%%%%%%%%%%%%%%%%%%%%%%%%
%%%%%%%%%%%%%%%%%%%%%%%%
\begin{figure}[t!]
\centering
\begin{tabular}{cc}    
    \subfigure[
    Differential cross-section $d\sigma/d\varphi$ as a function of $\varphi$ for $N = 1$ and $\alpha = 5/6$. The removal of an angular sector leads to an increased frequency of oscillations, indicating a more localized scattering effect.        
        \label{N1Plinear}
    ]{
        \includegraphics[width=70mm]{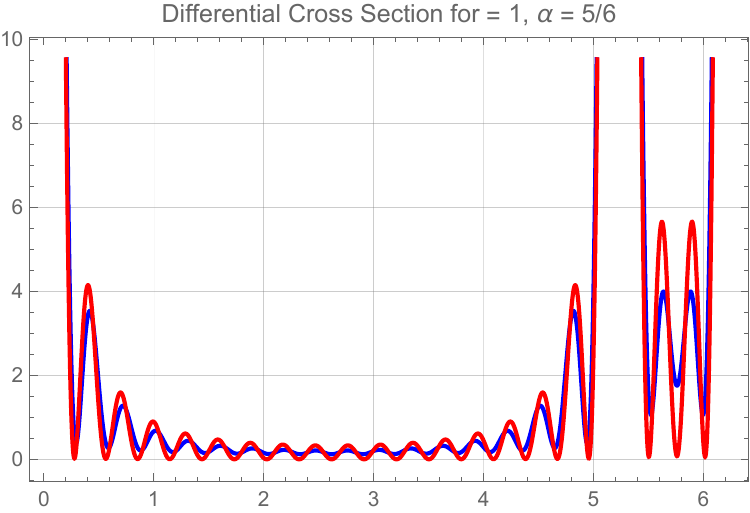}
    } &
    \subfigure[
        Polar plot of the differential cross-section for $N = 1$ and $\alpha = 5/6$. The confined angular range due to the missing sector enhances constructive interference in specific directions.
        \label{N1Ppolar}
    ]{
        \includegraphics[width=70mm]{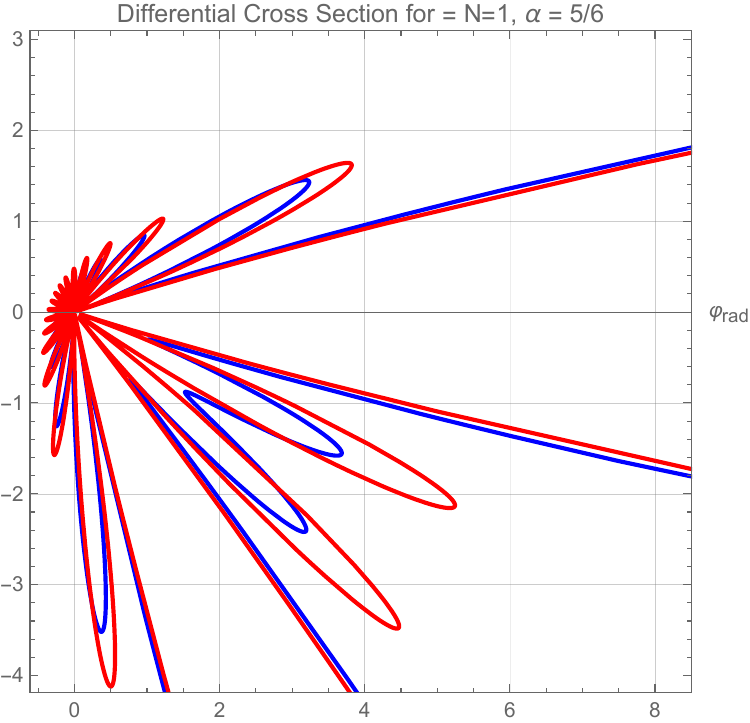}
    }
\end{tabular}
\caption{Differential cross-section $d\sigma/d\varphi$ for $N=1$ ($\alpha = 5/6$). 
The angular deficit confines the scattering space, leading to a higher frequency of oscillations in the linear plot and a more concentrated directional pattern in the polar representation.}

\label{N1P}
\end{figure}
%%%%%%%%%%%%%%%%%%%%%%%
%%%%%%%%%%%%%%%%%%%%%%%%
\begin{figure}[t!]
\centering
\begin{tabular}{cc}    
    \subfigure[
        Differential cross-section $d\sigma/d\varphi$ as a function of the scattering angle $\varphi$ for $N = -1$ and $\alpha = 7/6$. The scattering profile exhibits oscillations that reflect the topological deformation associated with the insertion of an angular sector, modifying the interference pattern
        \label{N1Nlinear}
    ]{
        \includegraphics[width=70mm]{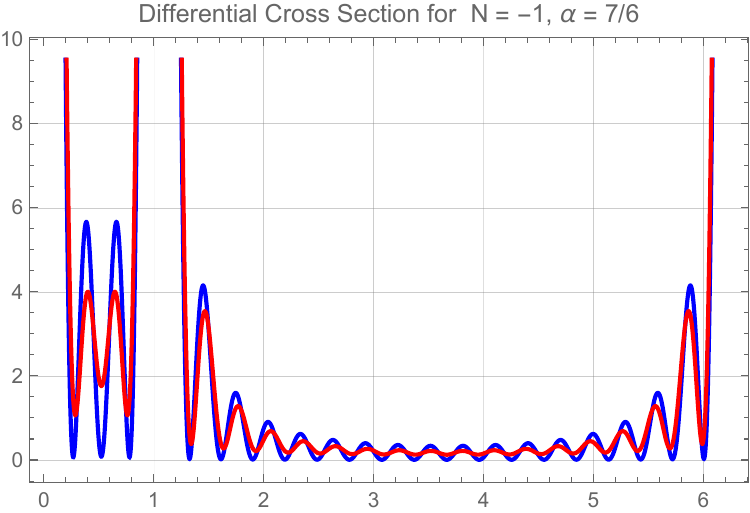}
    } &
    \subfigure[
        Polar representation of the differential cross-section for $N = -1$ and $\alpha = 7/6$. The modification of the angular range due to the inserted sector results in a redistribution of the scattering intensity over a broader range of angles.
        \label{N1Npolar}
    ]{
        \includegraphics[width=70mm]{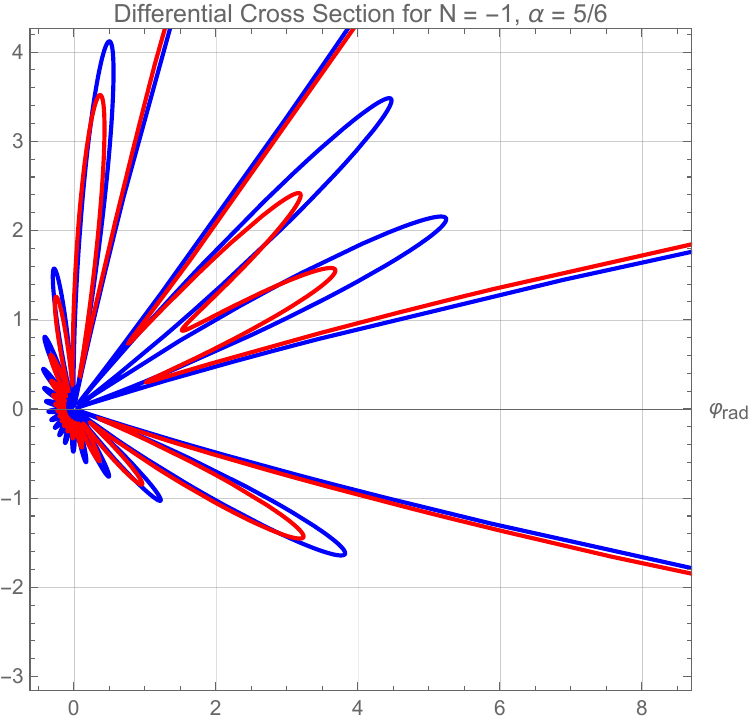}
    }
\end{tabular}
\caption{Differential cross-section $d\sigma/d\varphi$ for $N=-1$ ($\alpha = 7/6$). 
The angular excess expands the scattering space, resulting in more widely spaced oscillations in the linear plot and a more uniform scattering distribution in the polar representation.}

\label{N1N}
\end{figure}
%%%%%%%%%%%%%%%%%%%%%%%

The differential cross-section as a function of the scattering angle exhibits distinct behaviors for $N = 1$ and $N = -1$, as shown in Fig.~\ref{N1P} and Fig.~\ref{N1N}, respectively. These cases correspond to different topological defects in graphene, leading to variations in the interference structure due to the underlying geometric modifications.

For $N=1$ (Fig.~\ref{N1P}), which corresponds to a pentagon defect, the removal of an angular sector creates an angular deficit, reducing the available scattering space. This confinement leads to a higher frequency of oscillations in the linear representation (Fig.~\ref{N1Plinear}), where the peaks are sharp and closely spaced, indicating strong interference effects due to the restricted angular domain. In the polar representation (Fig.~\ref{N1Ppolar}), the scattering intensity is concentrated along specific directions, reinforcing the localized nature of the interference pattern.

In contrast, for $N=-1$ (Fig.~\ref{N1N}), which corresponds to a heptagon defect, an angular sector is added, creating an angular excess that expands the available scattering space. As a result, the oscillations in the linear representation (Fig.~\ref{N1Nlinear}) are more widely spaced, reflecting a lower interference density. The polar representation (Fig.~\ref{N1Npolar}) shows a more uniform scattering pattern, with intensity spread across a larger angular range, indicating reduced confinement effects.

The key difference between these two cases lies in how the angular modification influences the scattering response. In the angular deficit case ($N=1$), the confined space leads to high-frequency oscillations and a more localized scattering pattern. Conversely, in the angular excess case ($N=-1$), the expanded space results in a broader, more diffuse interference pattern with lower oscillation density. This contrast highlights the role of $\alpha$ in shaping the periodicity, intensity, and spatial distribution of the scattering process.

%%%%%%%%%%%%%%%%%%%%%%%%
\begin{figure}[t!]
\centering
\begin{tabular}{cc}    
    \subfigure[
        Differential cross-section $d\sigma/d\varphi$ as a function of $\varphi$ for $N = 2$ and $\alpha = 2/3$. The reduced angular range enhances the frequency of oscillations, indicating more localized interference effects.
        \label{N2Plinear}
    ]{
        \includegraphics[width=70mm]{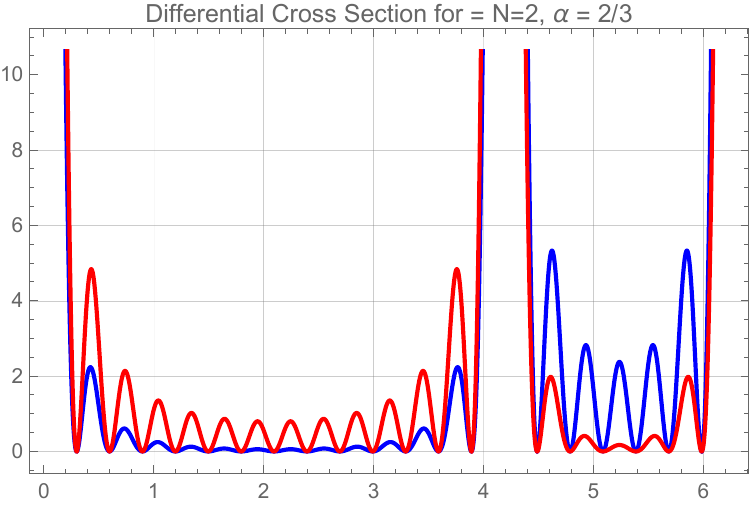}
    } &
    \subfigure[
        Polar plot of the differential cross-section for $N = 2$ and $\alpha = 2/3$. The reduction in available scattering angles leads to concentrated scattering directions with sharper intensity variations
        \label{N2Ppolar}
    ]{
        \includegraphics[width=70mm]{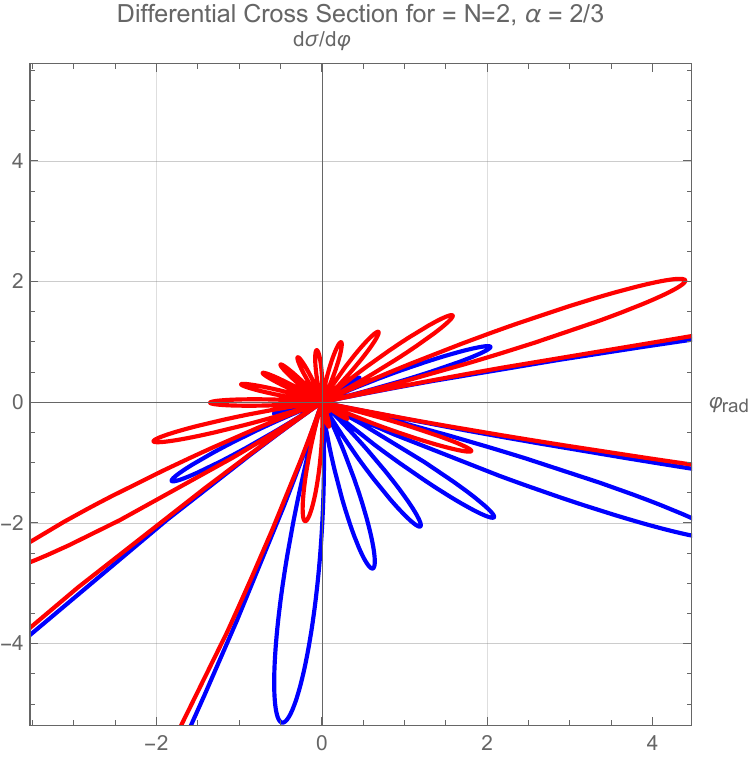}
    }
\end{tabular}
\caption{Effect of the parameter $ \Omega $ on the differential cross-section.}
\label{N2P}
\end{figure}
%%%%%%%%%%%%%%%%%%%%%%%
%%%%%%%%%%%%%%%%%%%%%%%%
\begin{figure}[t!]
\centering
\begin{tabular}{cc}    
    \subfigure[
        Differential cross-section $d\sigma/d\varphi$ as a function of $\varphi$ for $N = -2$ and $\alpha = 4/3$. The scattering pattern reflects the effect of an increased angular range due to the additional sector, decreasing interference density.
        \label{N2Nlinear}
    ]{
        \includegraphics[width=70mm]{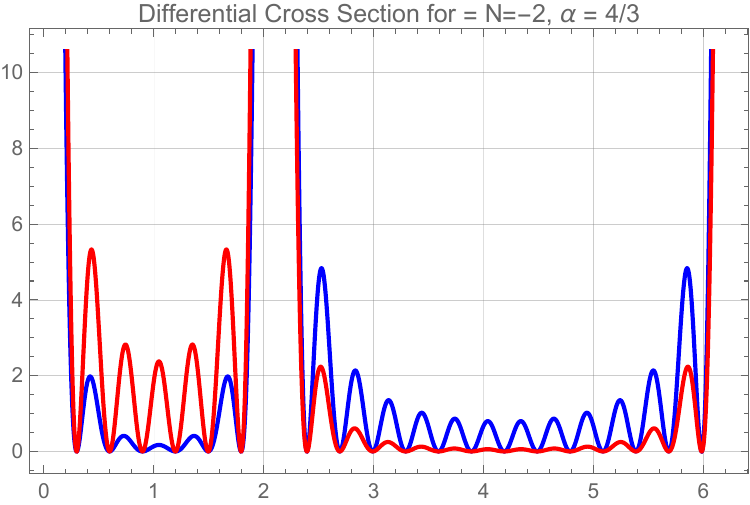}
    } &
    \subfigure[
        Polar representation of the differential cross-section for $N = -2$ and $\alpha = 4/3$. The broader angular domain disperses the scattering intensity over a larger range.
        \label{N2Npolar}
    ]{
        \includegraphics[width=70mm]{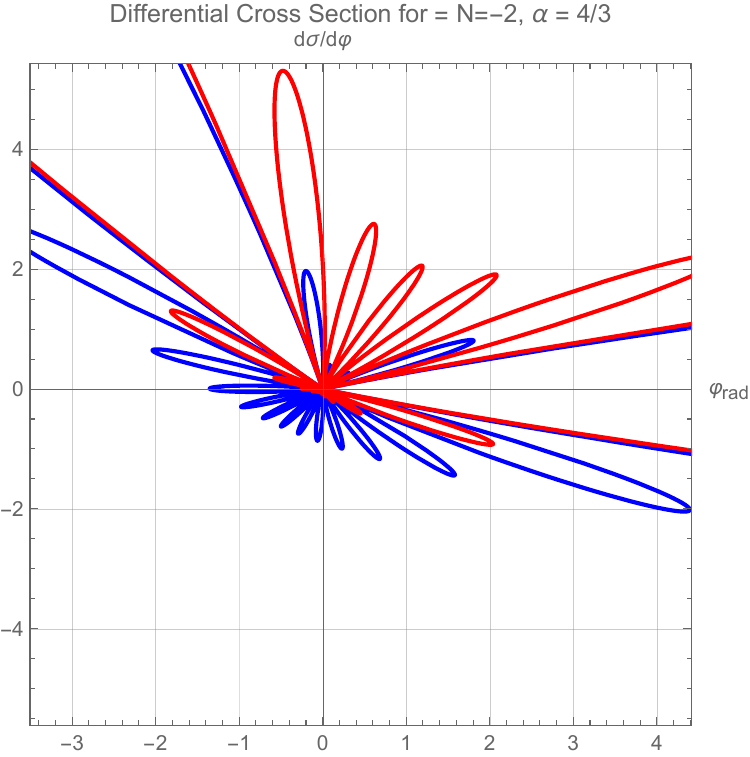}
    }
\end{tabular}
\caption{Effect of the parameter $ \Omega $ on the differential cross-section.}
\label{N2N}
\end{figure}
%%%%%%%%%%%%%%%%%%%%%%%%%%%
%%%%%%%%%%%%%%%%%%%
The differential cross-section as a function of the scattering angle also exhibits significant differences for $N = 2$ and $N = -2$, as shown in Fig.~\ref{N2P} and Fig.~\ref{N2N}, respectively. These cases correspond to distinct topological modifications in graphene, where the insertion or removal of angular sectors alters the interference structure of the scattered waves.

For $N=2$ (Fig.~\ref{N2P}), which corresponds to a square defect, the removal of two angular sectors leads to a stronger angular deficit, characterized by $\alpha = 2/3$. This reduction in available scattering space further enhances the confinement effect, increasing the frequency of oscillations in the differential cross-section. In the linear representation (Fig.~\ref{N2Plinear}), the peaks are even more densely packed compared to the $N=1$ case, indicating a highly localized interference structure. The polar representation (Fig.~\ref{N2Ppolar}) also reflects this behavior, showing that the scattering intensity remains concentrated along well-defined angular regions, reinforcing the directional nature of the interference.

For $N=-2$ (Fig.~\ref{N2N}), which corresponds to an octagon defect, the addition of two angular sectors produces an angular excess with $\alpha = 4/3$, significantly expanding the available scattering space. This geometric modification leads to a broader distribution of interference effects, as seen in the linear representation (Fig.~\ref{N2Nlinear}), where the oscillations are more widely spaced and less frequent. In the polar representation (Fig.~\ref{N2Npolar}), the scattering pattern appears more diffuse, with intensity distributed across a wider range of angles, reducing the directional confinement seen in the $N=2$ case.

Comparing the linear plots (Figs.~\ref{N2Plinear} and \ref{N2Nlinear}), we observe that the case $N=2$ exhibits a higher concentration of oscillations due to the restricted angular range, whereas for $N=-2$, the oscillations are more separated, reflecting the expanded space available for scattering. This highlights how the angular deficit leads to sharper, more frequent oscillations, while the angular excess results in a smoother interference profile.

A similar contrast is evident in the polar plots (Figs.~\ref{N2Ppolar} and \ref{N2Npolar}). For $N=2$, the scattering pattern is more directional, with well-defined peaks indicating stronger localization effects. In contrast, for $N=-2$, the scattering is spread over a broader angular range, with a more uniform intensity distribution. This comparison reinforces the role of $\alpha$ in shaping the scattering dynamics, where a reduced angular space confines the scattering directions, and an expanded space allows for a more diffuse and evenly distributed interference pattern.

%%%%%%%%%%%%%%%%%%%%%%%%
%%%%%%%%%%%%%%%%%%%%%%%%

%%%%%%%%%%%%%%%%%%%%%%%%%%%%%%%%%%%%%%%%%%%%%%%%%%%%%%%%%%%%%%%%%%%%%%%%%%%%%%%
%%%%%%%%%%%%%%%%%%%%%%%%%%%%%%%%%%%%%%%%%%%%%%%%%%%%%%%%%%%%%%%%%%%%%%%%%%%%
\begin{figure}[h]
    \centering
    \includegraphics[width=0.7\textwidth]{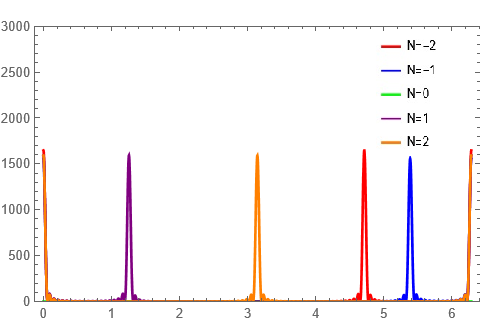}
   \caption{Combined Differential Cross-Section.}
   \label{fig:scatter_dsigma}
\end{figure}
%%%%%%%%%%%%%%%%%%%%%%%%%%%%%%%%%%%%%%%%%%%

%%%%%%%%%%%%%%%%%%%%%%%%%%%%%%%%%%%%%%%%%%%%%%%
In Fig.~(\ref{fig:scatter_dsigma}), the differential cross-section $d\sigma/d\phi$ is shown as a function of the scattering angle $\phi$ for different values of $N$. The figure provides a unified visualization of the scattering profiles previously analyzed, allowing a direct comparison of how the angular modifications introduced by disclinations influence the interference patterns. 

The results confirm that disclinations alter the scattering behavior by modifying the effective angular range. For $N > 0$, where an angular deficit is present, the confined space leads to a higher frequency of oscillations and more localized interference effects. In contrast, for $N < 0$, the insertion of angular sectors expands the scattering domain, reducing the density of oscillations and producing broader interference patterns. The transition from sharp, closely spaced oscillations for $N=2$ to smoother, more dispersed interference structures for $N=-2$ illustrates the progressive influence of increasing angular space on the scattering process.

This global comparison highlights the interplay between geometry and gauge effects in graphene with disclinations. The trends observed in Fig.~(\ref{fig:scatter_dsigma}) reinforce how the parameter $\alpha$ dictates the periodicity and spatial distribution of the interference patterns. The shifts in peak positions and amplitudes across different $N$ values provide insight into the underlying mechanisms governing quasiparticle scattering in curved graphene structures.

%%%%%%%%%%%%%%%%%%%%%%%%%%%%%%%%%%%
%%%%%%%%%%%%%%%%%%%%%%%%%%%%%%%%%%%%%%%%%%%%%%%%%%%%%%%%%%%%%%%%%%%%%%%%%%%%%%
\section{Summary and prospects}\label{sec6}

In this study, we conducted a comprehensive investigation of how disclinations affect quasiparticle scattering in graphene, emphasizing the intricate relationship between geometry, gauge fields, and electronic behavior. By extending the massless Dirac equation to account for curvature effects, we captured how topological defects alter electronic states and scattering profiles. The inclusion of a non-Abelian gauge field modeled sublattice mixing and shifts in the Dirac points, illustrating the gauge contribution's pivotal role.

Our findings demonstrate that disclinations produce distinct phase shifts and interference patterns, directly impacting the differential cross-section. We compared the angular deficit ($0 < \alpha < 1$) and angular excess ($\alpha > 1$) regimes, revealing contrasting scattering behaviors. The angular deficit regime compresses the angular range, resulting in high-frequency oscillations and localized interference patterns. Conversely, angular excess broadens the range, reducing interference density and producing more diffuse scattering profiles.

Additionally, we identified that the gauge field contribution, represented by $\Omega$, introduces a geometric phase similar to the Aharonov-Bohm effect, reinforcing the topological nature of the scattering modifications. A numerical case for $N=2$ was presented, illustrating how $\alpha$ and $\Omega$ jointly shape the scattering characteristics.

This study advances our understanding of the geometric and gauge influences on quasiparticle dynamics in graphene-like materials. Highlights how topological defects can be harnessed to control scattering properties, offering new avenues for defect-engineered materials. The insights provided here have implications for designing nanoscale electronic devices, quantum transport systems, and topologically robust materials that bridge fundamental physics with potential technological applications.

%%%%%%%%%%%%%%%%%%%%%%%%%%%%%%%%%%%%%%%%%%%%%%%%%%%%%%%%%%%%%%%%%%%%%%%%%%%%%

{\bf Acknowledgments:} This work was supported by Conselho Nacional de Desenvolvimento Cient\'{\i}fico e Tecnol\'{o}gico (CNPq) and Funda\c{c}\~ao de Apoio a Pesquisa do Estado da Para\'iba (Fapesq-PB). G. Q. Garcia would like to thank Fapesq-PB for financial support (Grant BLD-ADT-A2377/2024). The work by C. Furtado has been supported by the CNPq (project PQ Grant 1A No. 311781/2021-7).

\bibliographystyle{iopart-num}

  \bibliography{biblio} 

\end{document}